\documentclass[12pt]{iopart}

\usepackage{url,graphicx,calc,color}
\usepackage{iopams}
\newcommand{\csld}{\mbox{${\cal B}$}}
\begin{document}

\title[H-PDLCs for neutron optics]{Diffraction Gratings for Neutrons from Polymers and Holographic Polymer-Dispersed Liquid Crystals }

\author{M. Fally$^1$, M. Bichler$^1$, M. A. Ellabban$^{1,2}$, I. Dreven\v{s}ek Olenik$^3$, C. Pruner$^4$, H. Eckerlebe$^5$ and K.P. Pranzas$^5$}
\address{$^1$ Faculty of Physics, Nonlinear Physics, Vienna University, A-1090 Wien, Austria}
\address{$^2$ Physics Department, Faculty of Science, Tanta University, Tanta 31527, Egypt}
\address{$^3$ Faculty of Mathematics and Physics, University of Ljubljana and J.  Stefan Institute,  SI-1000 Ljubljana, Slovenia}
\address{$^4$  Faculty of Natural Sciences, Department of Material Science and Physics, University of Salzburg, A-5020 Salzburg, Austria}
\address{$^5$ GKSS Forschungszentrum, D-21502 Geesthacht, Germany}
\ead{martin.fally@univie.ac.at}

\begin{abstract}
We discuss the applicability of holographically recorded gratings in photopolymers and holographic polymer-dispersed liquid crystals as neutron optical elements.  An experimental investigation of  their properties  for light and neutrons with different grating spacings and grating thicknesses is performed. The angular dependencies of the diffraction efficiencies for those gratings are interpreted in terms of a rigourous coupled wave analysis.  Starting from the obtained results we work out the lines for the production of an optimised neutron optical diffraction grating, i.e., high diffraction efficiency in the Bragg diffraction regime with moderate angular selectivity.
\end{abstract}

\submitto{\JOA}
\maketitle
\renewcommand{\columnwidth}{76mm}
    \section{Introduction and statement of the problem\label{sec:intro}}
Neutron diffraction from light-induced gratings has proven to be of value for either, the material research aspect of gaining additional knowledge on photosensitive media\cite{Havermeyer-prl98,Havermeyer-phb00}, the technological aspect to improve such gratings for use as neutron optical elements \cite{Schellhorn-phb97}, as well as the aspect of fundamental physics \cite{Havermeyer-apb99,Rupp-osa99}. The major efforts in improving the quality of such gratings were made just for a single material, namely deuterated poly(methylmethacrylate) (d-PMMA)  and culminated in the successful setup of an interferometer for cold neutrons \cite{Pruner-nima06}. However, the gratings of which the interferometer is built up had to be aligned accurately once for ever during recording using an extraordinary demanding procedure because of their extreme angular selectivity. The latter is due to the large thickness $d$ of the gratings of about 2.8 mm which requires that the three gratings constituting the interferometer are aligned mutually parallel with an accuracy of at least $\mu$rad \cite{Pruner-nima06}.

Our goal is to reduce the technological problem of adjusting the gratings and to enhance the flexibility. Thus we are searching for alternative materials that match the following desired constraints, which constitute the 'optimal' diffraction grating suitable for a neutron interferometer:
\begin{enumerate}
\item The diffraction efficiency, defined as diffracted intensity over incident intensity,  of the grating should be as high as 50\%  for a beam-splitter and 100\%  for a mirror.
\item The diffraction takes place in the limit of two-wave coupling (Bragg diffraction regime)
\item The angular selectivity must not be too high, i.e., FWHM of the Bragg peak should not be less than 1/20 degree
\end{enumerate}
As will be shown below, a set of three parameters ($\lambda/\Lambda, d/\Lambda, \nu$) and their combination plays the main role in matching those constraints. Here, $\lambda$ is the neutron wavelength, $\Lambda$ the grating spacing, $d$ the grating thickness and $\nu$ the grating strength. For a pure phase grating the latter is defined as
\begin{equation}
\nu=\frac{\pi d \Delta n}{\lambda \cos{\theta_B}},
\end{equation}
where $\Delta n$ is the modulation of the refractive index and $\theta_B$ is the Bragg angle inside the medium.
It is obvious, that for such an alternative approach other materials than the standard d-PMMA have to be chosen. Therefore, we focus our interest on searching for novel materials for which the neutron-refractive index can be changed by light as well.

In this paper we show recent experimental results of neutron diffraction experiments from holographic polymer-dispersed liquid crystals (H-PDLCs) with various grating spacings, interpret them in terms of a rigourous coupled wave analysis and discuss by which means holographic polymer nanocomposites can be used for generating optimal diffraction gratings for cold neutrons.
    \section{Holographic polymer-dispersed liquid crystals}
Holographic polymer-dispersed liquid crystals (H-PDLC) have attracted attention because they combine the properties of liquid crystals, including their optical anisotropy and the fact that they are easily switchable in external electric fields, with that of  photopolymers. The latter allow for changes of the complex dielectric susceptibility upon illumination with light and therefore open up possibilities for various applications \cite{Sutherland-apl94,Escuti-apl00,Jakubiak-admat03,Escuti-ol03,Ren-apl03,Jakubiak-admat05,Ogiwara-ao08}.

 H-PDLCs are generated from photosensitive monomers which are fully miscible with liquid crystals. When a mixture is placed into a spatially modulated  light pattern, e.g.,  optical interference fringes, the monomers initially diffuse into the bright and the LC molecules to dark regions. After a certain degree of polymerization the components are unmixed and thus phase separated LC domains are formed. These LC domains comprise periodically modulated scattering centres for light which result  - besides a strong diffuse scattering background - in combined phase- and extinction gratings \cite{Drevensek-Olenik-pre06,Fally-oex08}. For neutrons the situation is quite different: the major contribution for diffraction stems from the density modulation between the LC and the polymer regions as well as from the difference in their mean coherent scattering lengths. Thus, we expect that for neutrons the gratings are of a pure phase-grating type as discussed below.
    \subsection{Preparation of the gratings: Holographic setup}
The samples were prepared from the mixture containing 55 wt.\% of the LC mixture (TL203, Merck), 33 wt.\% of the prepolymer (PN393, Nematel) and 12 wt\% of the 1,1,1,3,3,3,3-Hexafluoroisopropyl acrylate (HFIPA, Sigma-Aldrich) which is photosensitive in the ultraviolet spectral region \cite{Drevensek-Olenik-pre04}.  The cells to be filled with this mixture consist of two glass plates separated by spacers of $L=50\mu$m  thickness.
The gratings were fabricated by illuminating the samples in the interference field of two plane waves using an Argon ion laser operating at a wavelength of $\lambda_{UV}=351$ nm. The recording beams penetrated the four chosen samples symmetrically with respect to the sample normal at angles that resulted in grating spacings of $\Lambda_i=1200,1000,560,433$ nm, respectively. The total recording intensities of the s-polarized beams were  6 mW/cm$^2$ for the grating with spacing 1200 nm and 13 mW/cm$^2$ for the others with exposure times between 25 and 30 s. After the recording process the samples were postcured for about 6 min using a broadband uv-lamp  which was the optimum to get stable gratings according to our previous experience. The holographic two-wave mixing setup was stabilized using a feedback-loop system with an auxiliary grating similar to that described in Ref. \cite{De-Sio-ao06}, a technique which turned out to considerably assist the phase separation process and thus to improve the grating quality.

Problems during the recording process arise in particular for samples thicker than about 50 micron. The reason for this fact is obvious: after the growth of the LC droplets to sizes comparable to the light wavelength strong diffuse scattering occurs which completely destroys the spatially periodic interference pattern at larger sample depths and even results in nonlinear small signal amplification effects, i.e., holographic scattering \cite{Ellabban-apl05}. A possibility to overcome this obstacle could be to record the grating at high electric fields which suppresses holographic scattering \cite{Ellabban-om07} or to heat the sample above the nematic-isotropic phase transition. However, as a matter of fact we are basically limited to thicknesses of several tens of microns.
\begin{table}
\caption{\label{tab:samples} Summary of the grating preparation parameters for the investigated samples.}
\centering
\begin{tabular}{ccrc}\hline
Sample & L [$\mu$m]& \multicolumn{1}{c}{$\Lambda$ [nm] }&Exposure [$\rm Ws/cm^2$]\\\hline
10f & 50 & 1200 & 210\\
29b & 50 &  1000 & 390\\
29w & 50 & 560 & 312 \\
29y & 50 & 433 & 390\\\hline
\end{tabular}
\end{table}
    \subsection{Neutron optics with H-PDLCs: Short course}
An exact description of neutron optics is given by the one-body Schr\"odinger equation with a neutron optical potential  and deals with coherent elastic scattering phenomena in condensed matter such as diffraction, reflection and refraction \cite{Sears-89}. Neutron optics is reasonable with low energy neutrons, i.e., thermal, cold or ultra-cold neutrons, which corresponds to wavelengths in the range $10^{-1} - 10$ nm. Here, we use cold neutrons of about 1-3 nm wavelength.  For our concerns the dominating term of the neutron-optical potential is the Fermi pseudo-potential, i.e., the nucleus-neutron interaction, which yields for the macroscopic neutron-optical potential
\begin{equation}
V(\vec{r})=\frac{2\pi\hbar^2}{m}b\rho(\vec{r}).
\end{equation}
Here, $m$ is the neutron rest mass, $b$ the mean bound coherent scattering length and $\rho(\vec{r})$ the number density. As in light optics the decisive quantity in neutron optics is the refractive index. By a comparison of the Helmholtz equations derived from either the Schr\"odinger equation or Maxwell's equations, the refractive index for neutrons is defined as:
\begin{equation}
n=\sqrt{1-\frac{V}{E}}
\end{equation}
with $E$ the total energy.
We are now interested in the modulation of this neutron-refractive index after illumination with a sinusoidal light-interference pattern. Assuming that the light pattern is transformed linearly into a light-induced modulation of the neutron-refractive index, the latter reads \cite{Fally-apb02}:
\begin{equation}\label{eq:deltan}
\Delta n(\vec{r})=\frac{\lambda^2}{2\pi}{\Delta\cal B}(\vec{r})
\end{equation}
where ${\cal B}=b\rho$ is the so-called bound coherent  scattering length density.  Assuming that one would be able to periodically separate the polymer and the LC components completely, i.e.,  into slices of polymer and LC \cite{Caputo-ol04}, we would expect a  modulation of the neutron refractive index proportional to $\Delta{\cal B}=|{\cal B}_{LC}-{\cal B}_{P}|$.
	\section{Experimental results}
Our investigations were focussed on four samples with different grating spacings from two series of recording. The sample characteristics are summarized in table \ref{tab:samples}. The samples from series 29 were produced just one month before the neutron diffraction measurements (04-2008) whereas sample 10f was made about 6 months prior to the neutron experiments (07-2005). As ageing effects and dark polymerization are known to be prominent in these acrylate compounds \cite{Ellabban-apb08}, a quantitative comparison between the two series of the samples might be misleading.  The samples were characterized by using light diffraction experiments, polarization microscopy as well as atomic force microscopy (AFM), the results of which will be shown here only for one example. We will focus on the neutron diffraction experiments, their interpretation and perspectives.

\begin{figure}\centering
  \includegraphics[width=\columnwidth]{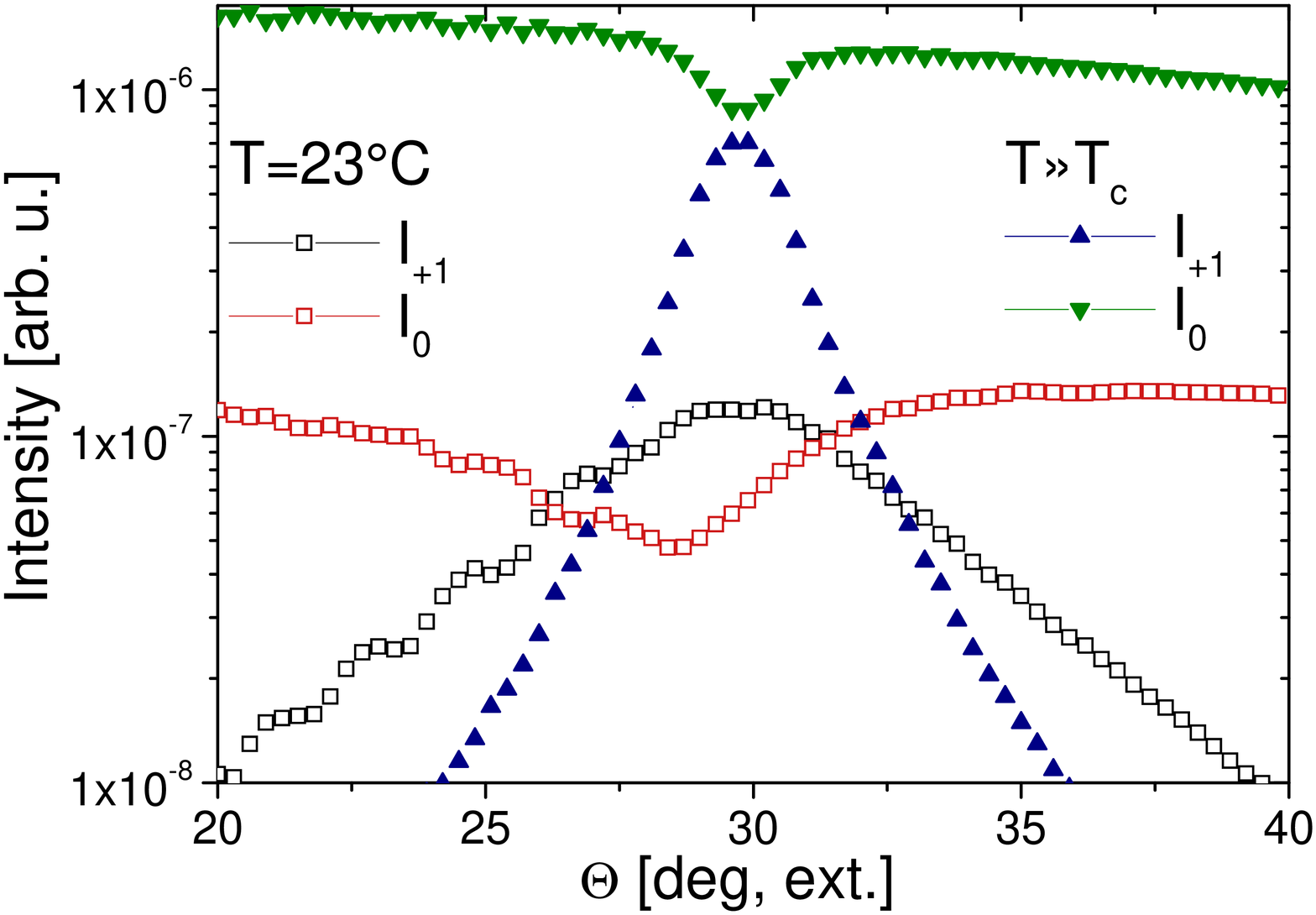}
  \caption{First and zero order light diffraction efficiencies for sample 29w at room temperature and  above the nematic-isotropic phase transition of the LC material. P-polarized  light at a wavelength of 543 nm was used. Note, that at room temperature considerable extinction effects can be observed and that the grating is obviously strongly overmodulated \cite{Fally-oex08}. These features are absent at high temperatures when the LC component is in the isotropic phase.}\label{fig:29w543}
\end{figure}

The light optical properties of the sample from series 10,  in particular the temperature dependence of the light diffraction, were published in a previous paper \cite{Drevensek-Olenik-pre06}. For series 29 of the samples we conducted measurements of the angular dependence of the diffraction efficiencies (rocking curves) at ambient temperature as well as above the nematic-isotropic phase transition. The efficiency $\eta_s$ of the $s$-th diffraction order is defined as
\begin{equation}\label{eq:diffeff}
\eta_s=\frac{I_s}{I}
\end{equation}
where $I$ is the intensity of the incident beam and $I_s$ the intensity of the $s$-th order diffracted beam.
The rocking curves in the isotropic phase were used for determining the thickness of the gratings, an issue which is not as simple as expected. At room temperature the diffuse scattering effects and the extremely large grating strength $\nu$ (overmodulation) for light diffraction lead to a complicated structure of the angular dependence for the diffracted intensities and hence a quantitative analysis is difficult. As an example we show the rocking curves in Fig. \ref{fig:29w543}, i.e., first and zero order diffraction,  for sample 29w using $p$-polarised light ($\lambda=543$ nm) at  room temperature and well above the phase transition temperature. It is evident that the extinction is extremely high in the low temperature phase which results in a more than one order of magnitude lower diffraction intensity.

\begin{figure}\centering
\includegraphics[width=\columnwidth]{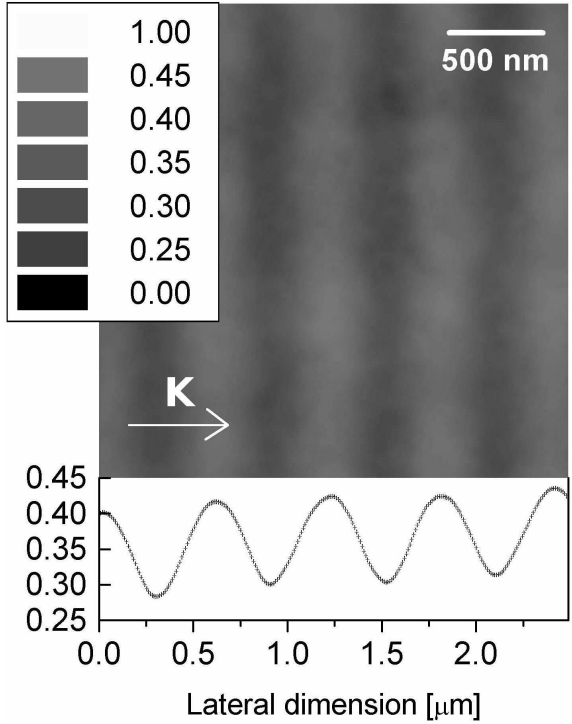}
\caption{\label{fig:afm}
AFM picture of the polymer matrix of sample 29w. The full size of the picture scales to $2.5\times 2.5\mu$m$^2$. Lower graph is  a cross-section obtained by averaging perpendicular to the grating vector.}
\end{figure}

In Fig. \ref{fig:afm} an AFM picture of the polymer matrix of sample 29w is shown. Before AFM imaging one glass plate was detached and the LC material was washed out from the grating structure with isopropanole.  The grating spacing from the AFM is estimated by a fast Fourier transform to be $\Lambda=556$ nm in good agreement with the expected value of 560 nm. Despite the fact that the profile (lower graph) seems to be nicely sinusoidal, enhanced values of some higher Fourier coefficients show, that the polymer and  LC regions are quite well separated, i.e., a sliced structure (POLICRYPS) is obtained \cite{Caputo-ol04,Caputo-josab04}. This fact might be attributed to the more sophisticated setup (stabilization) as compared to our previous recording experiments \cite{De-Sio-ao06}.

The neutron diffraction experiments were conducted at the Geesthacht neutron facility (GeNF) employing the small angle scattering instrument SANS-2. Full collimation of 20 m with a rectangular aperture of $10\times 40$ mm$^2$ at the entrance and a circular one of 5 mm  diameter was chosen to ensure sufficiently low angular spread of the incident neutron beam. The sample-detector distance was also at its maximum value of about 21 m owing to the small diffraction angle $\lambda/\Lambda\approx 10^{-3}$ rad. The detector matrix has a resolution of $256\times256$ pixel, each of $2.2\times 2.2$ mm$^2$ area. Basically, two different mean wavelengths of $\lambda=1.16$ nm and $\lambda=1.96$ nm with a spread of 10\% were employed.
The samples were placed on a rotation stage with a resolution of 1/1000 degree to measure the angular dependence of the diffraction efficiencies.
At each angular point we collected about $10^5$ total counts in 20 minutes which limits the relative error for the diffracted beam intensity to less than 10\%.
For an accurate evaluation an off-Bragg measurement was subtracted as background.  The scattering background of  the 1mm thick glass plates (empty cells) yielded a transmission of about 97\%.

\begin{figure}\centering
\includegraphics[width=\columnwidth]{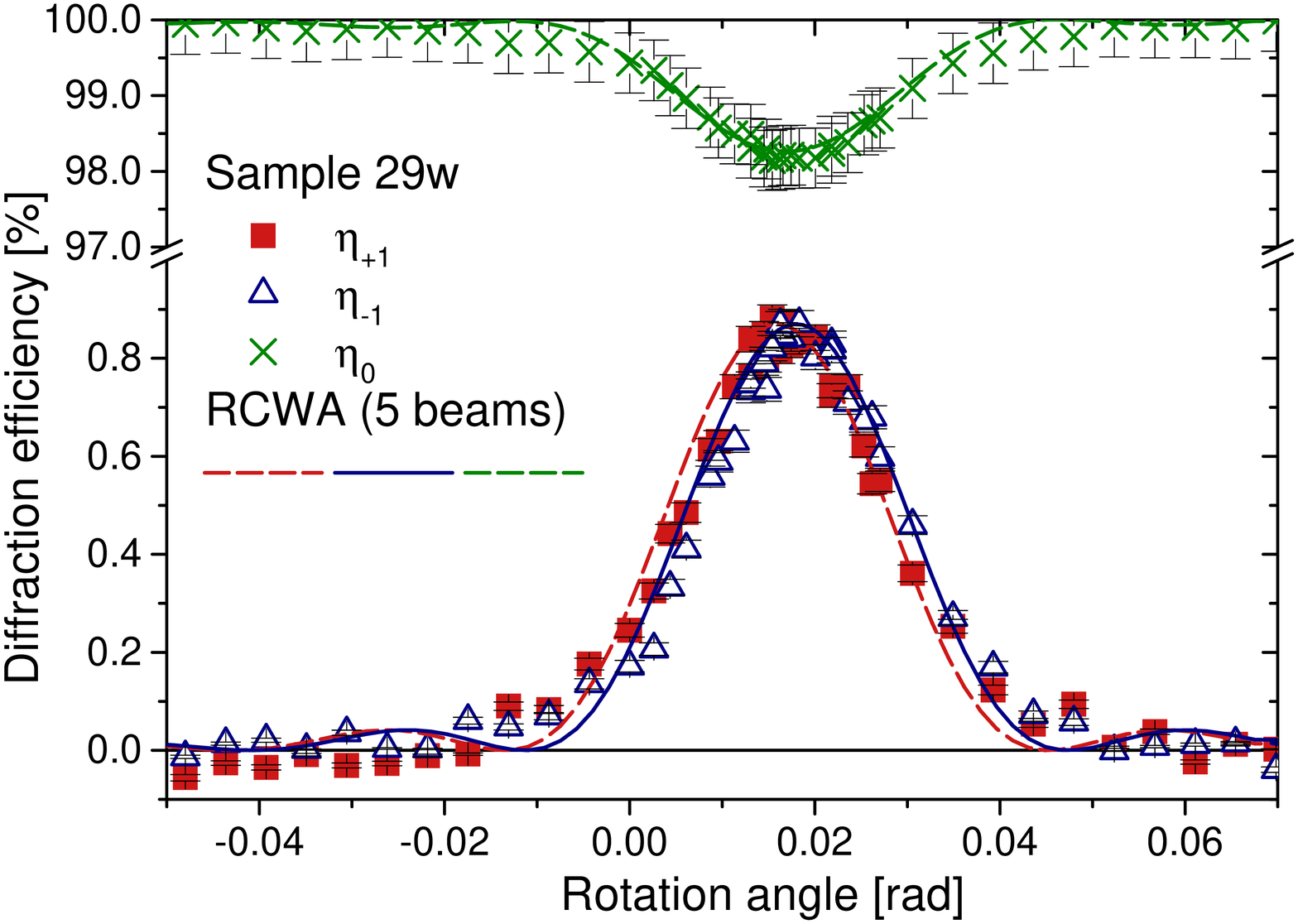}
\caption{\label{fig:rockn} Angular dependence of the diffraction efficiency for the zero and $\pm$ first orders for neutrons at a wavelength $\lambda=1.16$ nm of sample 29w. The solid lines show results of the RCWA with the values of table \ref{tab:sampres}. Five-wave coupling was largely sufficient, as diffraction orders higher than the first ones had less than 1/1000\% efficiency.} 
\end{figure}

In Fig. \ref{fig:rockn} the rocking curves for the zero and first diffraction orders for sample 29w are shown as a typical example.
The diffraction efficiency at the Bragg angle amounts only to about 0.9\% which is much lower as compared to sample 10f \cite{Drevensek-Olenik-spie07}.
Due to the very small values of $\lambda/\Lambda$ and $d/\Lambda$ at least three coupled waves are propagating: the zero order together with $\pm 1$  orders, i.e., evidently we are not in the Bragg diffraction regime. On the other hand one can notice a moderate but non-negligible angular selectivity, i.e.,  a dephasing occurs for off-Bragg positions which is neglected in the Raman-Nath theory applicable for the description of thin gratings \cite{Gaylord-apb82}. In general it would therefore be necessary to treat the results in the frame of a rigourous approach (mode coupling or wave coupling=RCWA) to fully describe the observed diffraction properties \cite{Moharam-josa81}.  This was done for sample 29w and the results of the RCWA analysis for sample 29w are shown as solid lines in Fig. \ref{fig:rockn}.
Fortunately, however, for such low diffraction efficiencies of $\eta<5\%$ with grating strengths $\nu<1/4$,  the first order diffraction efficiencies derived from the Raman-Nath, the two-wave coupling, and the RCWA theories lead to the same result within an accuracy of 10\%:
\begin{equation}\label{eq:etasimple}
\eta_{\pm 1}(\theta_B)\approx\nu^2
\end{equation}
Nevertheless, measuring rocking curves is important because they enable us to subtract the background properly and to evaluate the effective grating thickness $d$. We found that  in H-PDLCs $d$ is typically much less than the spacer's distance $L$  (see table \ref{tab:sampres}).

\begin{figure}\centering
\includegraphics[width=\columnwidth]{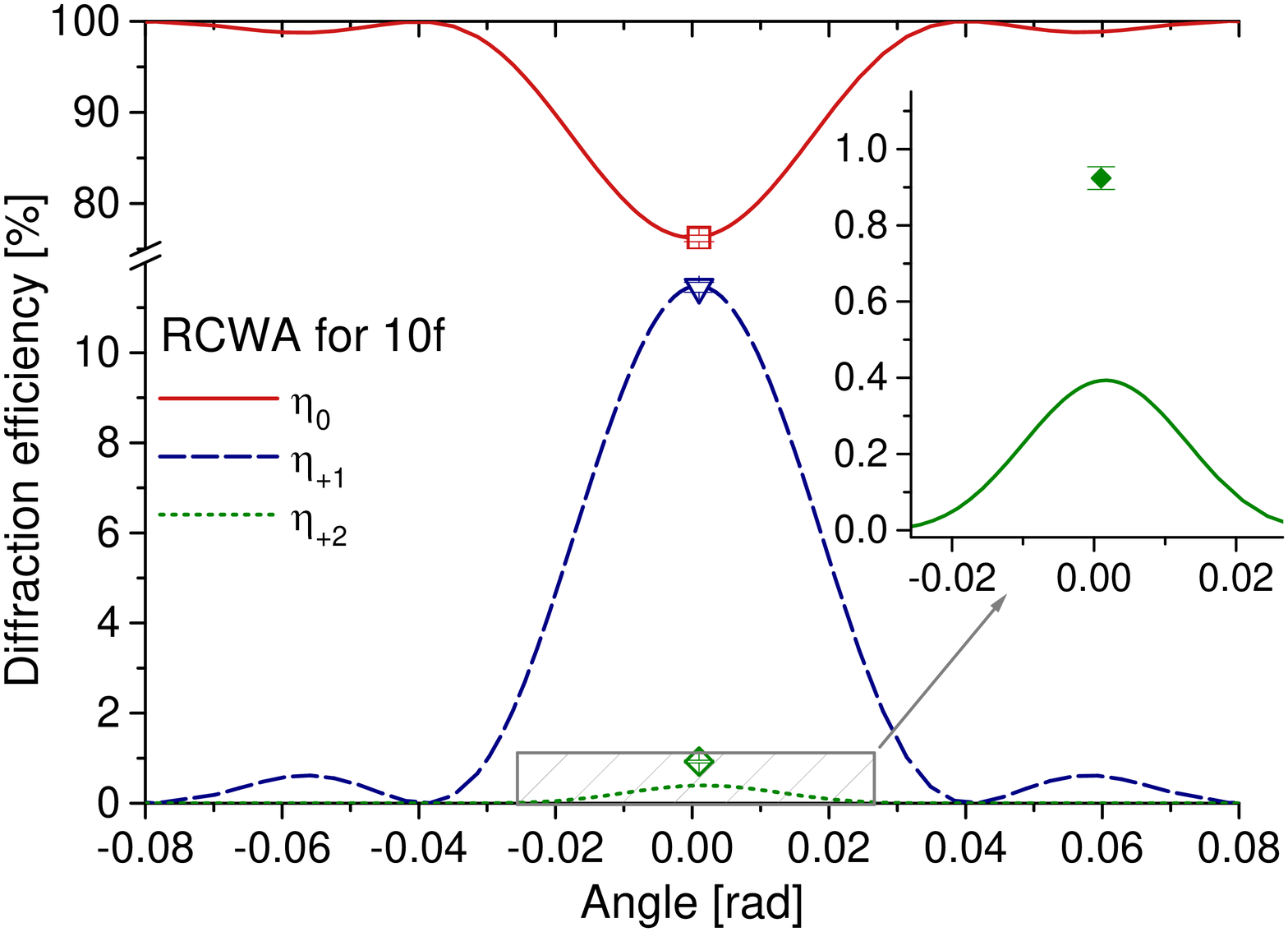}
\caption{\label{fig:rcwa} Rigourous coupled wave analysis for sample 10f.  We employed seven-beam coupling with diffraction efficiencies of the third orders less than 1/100\%.  Inset: Zoom in of the second order diffraction efficiency.}
\end{figure}

The RCWA approach, however, remains important as demonstrated for the case of sample 10f  measured at the long wavelength of $\lambda=1.96$ nm.  As can be seen from (\ref{eq:nu}) the refractive-index modulation for neutrons is proportional to $\lambda^2$. Thus, for the same grating an increase of $\lambda$ causes a profound increase of the grating strength $\nu$. As will be discussed in section \ref{sec:regime} (Fig. \ref{fig:expregime})  this might lead to a different diffraction regime. For $\lambda=1.96$ nm the first order diffraction efficiencies $\eta_{\pm 1}$ of sample 10f are about 10\% and also second order diffraction peaks show up \cite{Drevensek-Olenik-spie07}.  By illuminating the samples with a sinusoidal fringe pattern as described above, we expect also a sinusoidal shape of the refractive index modulation. If higher harmonics appear they might stem either from a deformed shape of the pattern due to strong nonlinearities of the recording process or just from the fact that diffraction occurs in the so called thin diffraction regime (Raman-Nath regime, see next section) with multiwave coupling. To discriminate between these two effects we again employ the RCWA for sample 10f. The result is shown in Fig. \ref{fig:rcwa} together with the experimental points of all diffracted orders ($0, +1, +2$) near Bragg  incidence.
  The measured diffraction efficiency for the second order is more than a factor of 2 larger than expected from the rigourous coupled wave analysis, which fits the values for the zero and first orders well. Therefore, we draw the conclusion that the grating structure indeed contains higher harmonics.
Finally, we calculated the coherent scattering length density modulation $\Delta{\cal B}$ for all investigated samples by employing (\ref{eq:etasimple}) and (\ref{eq:nu}). The results are summarized in table \ref{tab:sampres}.  One can notice a tendency that for a smaller grating spacing the refractive-index modulation and thus $\Delta\csld$ is reduced. This seems to be a general property of acrylate compounds as similar effects were observed in polymer dispersed nanoparticle systems \cite{Suzuki-apl02} and d-PMMA \cite{Rupp}.
\begin{table}\centering
\caption{\label{tab:sampres} Sample properties and parameters for $\lambda=1.19$ nm. The grating thickness $d$ was determined either by the neutron rocking curve ($d_n$), the light diffraction experiments in the isotropic phase ($d_o$) or by polarization microscopic means ($d_m$) after detaching the glass plates and washing out the LC  components.}
\begin{tabular}{cccc|cccc}\hline
Sample & $d_n$ &$ d_o$ &$ d_m$  &$\Delta{\cal B}$ [$\mu{\rm m}^{-2}] $&$\lambda/\Lambda$&$d/\Lambda$& $\nu$\\\hline
10f &  30 & 30  & -  &9.9 & $0.97\times 10^{-3}$ & 25 & 0.17\\
29b&  -  & 39  & 37  &5.1 & $1.2\times 10^{-3}$ &38 & 0.12\\
29w & 20  &  31  &  31 &8.2  & $2.1\times 10^{-3}$ &36 & 0.09 \\
29y & -  & 29  & 29  & 1.7  & $2.7\times 10^{-3}$ & 67 & 0.03 \\\hline
\end{tabular}
\end{table}

\begin{figure}\centering
\includegraphics[width=\columnwidth]{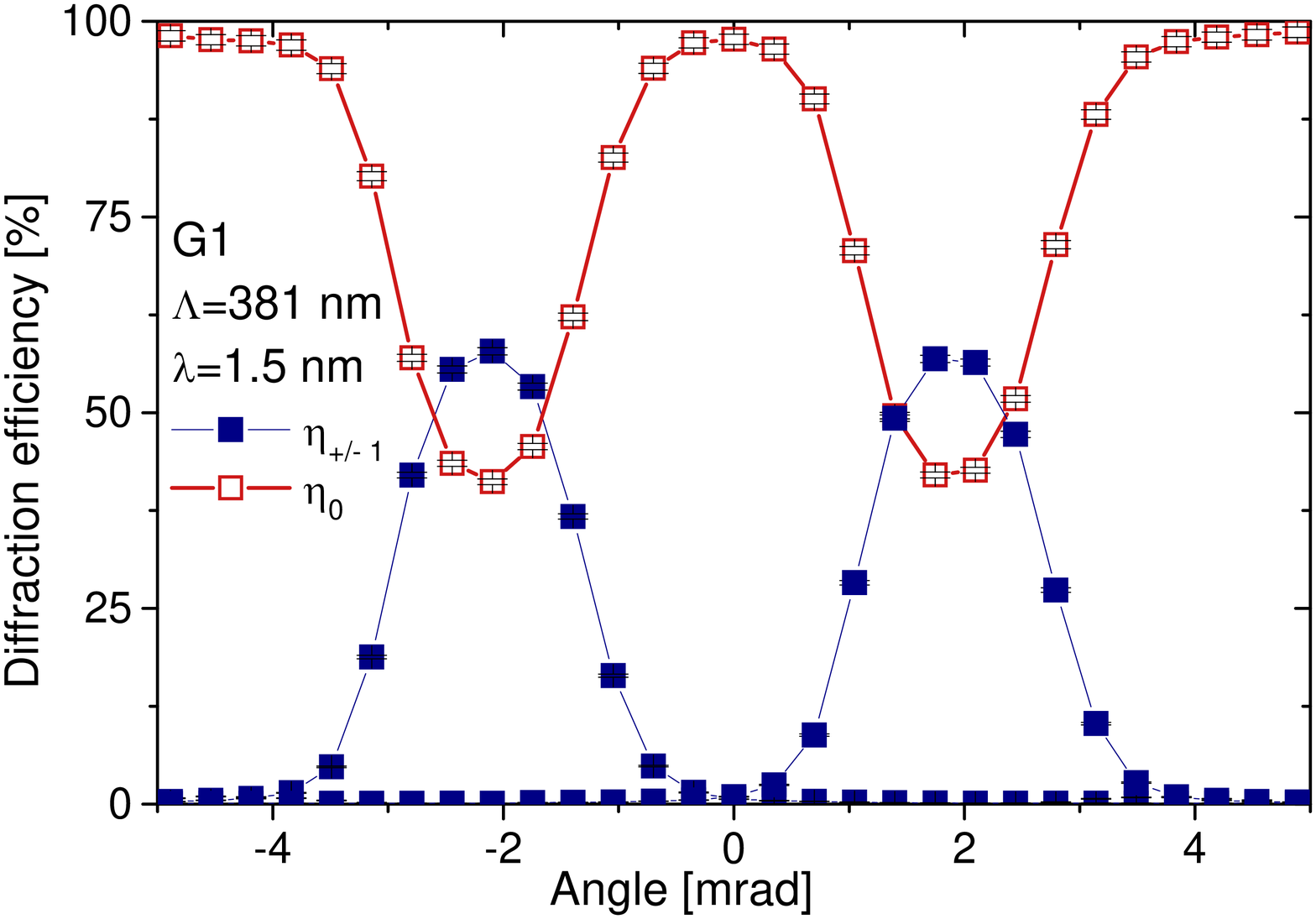}
\caption{\label{fig:pmmaG1} First and zero order diffraction efficiencies for a grating recorded in d-PMMA: $\Lambda=381$ nm, $d=2.7$ mm and the neutron wavelength is $\lambda=1.5$ nm.}
\end{figure}

Deuterated poly(methylmethacrylate) was and still is the most extensively used material to produce light-induced gratings for neutron diffraction \cite{Schellhorn-phb97,Rupp-prl90,Havermeyer-00}. One of its drawbacks is the uncontrollable dark polymerisation which lasts for weeks and further develops the grating. Another problem results from the large  sample thickness of 2-3 mm. Due to this large thickness, the grating's angular selectivity - given by the parameter $\Lambda/d$ -  is extremely high ($\sim 1/100$ degree),  so that rocking curves rather reveal the shape of the angular distribution of the neutron beam than that of the actual grating. This makes the evaluation of the data more complicated and ambiguous, because a deconvolution of the rocking curve with the angular and wavelength distribution has to be performed. Partially these problems were solved in Ref. \cite{Havermeyer-phb00,Havermeyer-00} employing certain approximations. To demonstrate the problem we show in Fig. \ref{fig:pmmaG1} a rocking curve for a grating with a thickness of about 2.7 mm and a grating spacing $\Lambda=381$ nm measured by cold neutrons at a wavelength of 1.5 nm.
Even if the rocking curves look quite nice at first glance (and they are nice!), the evaluation of the grating strength cannot be deduced directly from the maximum value of the diffraction efficiency as we must take into account the angular and wavelength distribution of the neutron beam and possible overmodulation effects of the grating. To overcome this problem wavelength dependence of the diffraction efficiency was measured \cite{Pruner-nima06,Pruner-04} and this lets us deduce the modulation of the coherent scattering length density $\Delta{\cal B}$ of about $(9\pm 1) \mu {\rm m}^{-2}$ which is pretty much the same as for the H-PDLC sample 10f \cite{Drevensek-Olenik-spie07,Fally-prl06}.
However, contrary to d-PMMA, the HPDLC mixture used in our study was not at all optimized for neutron diffraction and does not include any deuterated compounds.
The set of parameters for d-PMMA then is: $\lambda/\Lambda=4\times 10^{-3}, d/\Lambda=7\times 10^3, \nu=18$. Obviously, the latter two parameters differ largely from those of H-PDLCs, mainly due to the difference in the geometrical grating thickness.
    \section{Optimised diffraction grating\label{sec:regime}}
Let us now quantify the requirements for an optimised diffraction grating as addressed in section \ref{sec:intro}. The grating strength $\nu$ for neutrons  using (\ref{eq:deltan}) is:
\begin{equation}\label{eq:nu}
\nu=\frac{1}{2}d \lambda \Delta{\cal B}.
\end{equation}
Thus the grating strength depends linearly on the thickness, the wavelength and the coherent scattering length density modulation.
To ensure the Bragg diffraction regime two criteria must be fulfilled at the same time \cite{Gaylord-ao81}:
\begin{eqnarray}\label{eq:moharam1}
Q\nu&>&1\\
\frac{Q}{2\nu}&>&10\label{eq:moharam2}
\end{eqnarray}
with the famous parameter $Q=2\pi\lambda d/(n \Lambda^2 \cos{\theta_B})$ and $n$ the refractive index.  Note, that it is not sufficient just to fulfill one of these inequalities. These two inequalities define a region of permitted combinations $Q, \nu$ for which two wave diffraction occurs.
The angular selectivity is directly related to the parameter $d/\Lambda$.
 Just to introduce a measure for the accepted minimum width of the rocking curve we decide to say that the FWHM should be 1/20 of a degree or more, so that
\begin{equation}\label{eq:doverlambda}
d/\Lambda<600\sqrt{1-(\nu/\pi)^2}.
\end{equation}
By logarithmising equations (\ref{eq:moharam1})  and (\ref{eq:moharam2})  two linear inequalities in $\log{(Q)}, \log{(\nu)}$ result. The corresponding equalities geometrically yield two straight lines splitting the space into three regimes: both inequalities are fulfilled (Bragg diffraction regime), both are violated (Raman-Nath diffraction regime) and only one of them is violated/fulfilled. For the latter regime the use of the RCWA  is mandatory.

\begin{figure}\centering
\includegraphics[width=\columnwidth]{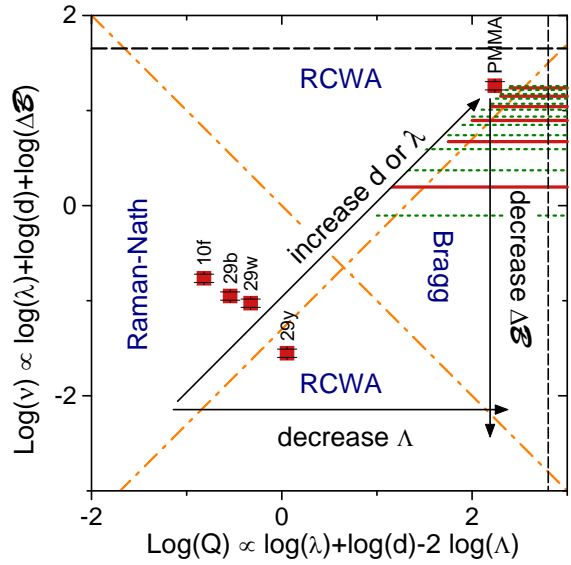}
\caption{\label{fig:expregime} Diffraction regimes separated by orange dash-dotted lines and experimental data. Regions of high diffraction efficiency for mirrors or beam-splitters are shown as red solid and green dotted lines, respectively. The arrows indicate possible paths to move between diffraction regimes by changing the corresponding parameters.}
\end{figure}

On the basis of the obtained neutron diffraction experiments the limits for the decisive parameters $d/\Lambda,Q, \nu$ form experimental and/or material constraints:
\begin{eqnarray}\nonumber
10^{-1}<&d/\Lambda&<10^4 \\
10^{-5}<&Q&<10^{3}\nonumber\\
&\nu&<45. \nonumber
\end{eqnarray}
Before addressing the final goal giving the set of parameters for an optimal diffraction grating let us at first have a look on the experimental values obtained for the H-PDLCs and the d-PMMA sample. In Fig. \ref{fig:expregime} $\log(\nu)$ is plotted  as a function of $\log(Q)$ for the investigated samples.  Also shown are the borders between the diffraction regimes and the regions of high diffraction efficiency (mirror indicated as red solid line, beam-splitters in green dotted). Here, we also included the case of overmodulated gratings, i.e., $\nu>\pi/2$.
We draw the attention to the fact that \emph{none} of the samples  completely fulfills the requirements for an optimal diffraction grating. The H-PDLC samples operate in the Raman-Nath regime, whereas the d-PMMA sample has much too high angular selectivity and diffraction occurs in the RCWA regime. The latter is astonishing at first glance because the sample is thick and the grating strength extremely high. By inspecting Fig. \ref{fig:expregime} it becomes clear, that the coherent scattering length density is \emph{too high} to meet the Bragg regime (for given values of $\Lambda, d$). Further we learn, that for the H-PDLC samples 10f, 29b, 29w an increase of the wavelength - the only parameter we can vary after production of the grating -  (or thickness) never will result in a two-wave coupling diffraction regime. For sample 10y, however, we could simply drive to the Bragg diffraction regime by increasing the wavelength. The reason is twofold: small grating spacing and a relatively low coherent scattering length density. This sounds somehow like a paradox but it is a matter of fact and just a consequence of the  conditions (\ref{eq:moharam1}) and (\ref{eq:moharam2}) to be simultaneously fulfilled.  Yet we should keep in mind that we would like to produce mirrors and beam-splitters. This requires high diffraction efficiencies and according to (\ref{eq:nu}) a low $\Delta\cal{B}$ then just can be compensated by  $\lambda$ and/or $d$. Further we made a demand on having moderate angular selectivity in (\ref{eq:doverlambda}) which basically can only be fulfilled for $d$ in the hundred micrometre range. To summarize: grating thickness and modulation of the coherent scattering length density should be kept moderate, the grating spacing as low as experimentally viable and the wavelength as large as possible. Indeed, the range of possible parameters is very narrow.

\begin{figure}\centering
\includegraphics[width=\columnwidth]{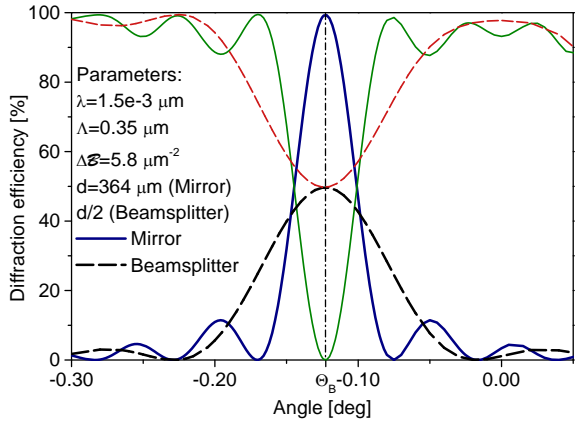}
\caption{\label{fig:optimal} First order and zero order diffraction efficiencies for an optimized grating. Solid lines give the angular dependence of the reflectivity for a mirror, dashed lines for a 50/50 beam-splitter. $\lambda=1.5$ nm, $\Lambda=350$ nm, $\Delta{\cal B}=10/\sqrt{3} \mu{\rm m}^{-2}, d=210\sqrt{3}\mu$m. Grating thickness for the beam-splitter is halved.}
\end{figure}

The set of parameters for a mirror and a beam-splitter with moderate angular selectivity, manageable values of  the grating spacing $\Lambda=350$ nm and reasonable values for $\Delta{\cal B}$ at cold neutron wavelengths require a grating thickness of about 200 $\mu$m. This value cannot easily be reached with H-PDLCs because of diffuse and holographic scattering \cite{Ellabban-apl05}. For d-PMMA this thickness should be doable and  $\Delta{\cal B}$ might be controllable by reducing the modulation depth $m=2\sqrt{I_RI_S}/(I_R+I_S)$ of the light pattern, where $I_R,I_S$ are the intensities of the interfering beams. However, currently this is  just an interesting speculation.
Therefore, we suggest to fabricate the optimal grating for use as mirrors or beam-splitters from holographic polymer-dispersed nanoparticles as introduced by \emph{ Suzuki and Tomita} \cite{Suzuki-apl02,Suzuki-ao04,Tomita-ol05,Suzuki-ao07}. Coherent scattering length densities can then be tailored by using the proper nanoparticles and a grating thickness  of 200 $\mu$m was already obtained \cite{Suzuki-ao07}. To conclude the discussion, in Fig. \ref{fig:optimal} we show the angular dependence of the diffraction efficiencies for a mirror and a 50/50 beam-splitter  employing optimal diffraction gratings.
With the feasible parameters chosen for the simulation it is possible to reach our initially formulated goal. The inequalities (\ref{eq:moharam1}) are almost fulfilled so that only two waves are propagating with considerable intensities.
For the case of the mirror $\nu=\pi/2$, so that the angular selectivity according to (\ref{eq:doverlambda})  is about 1/10 of a degree. Using the RCWA we find, that diffraction efficiencies for others than the zero and first order at the Bragg angle are less than 0.5\%. Thus, we can fairly say that a mirror for cold neutrons can be fabricated within a narrow range of parameters $d,\Lambda,\Delta{\cal B}, \lambda$ by means of optical holography in photo neutron-refractive polymers and polymer nanocomposites.
    \section{Discussion and Summary}
As discussed in the previous sections, optimised light-induced diffraction gratings for neutrons can be recorded only in a narrow range of parameters, of which $\Delta{\cal B},d,\Lambda$  partially depend on the choice of available materials. The advantage of d-PMMA is that we have rich experience in the preparation, and the modulation of $\Delta{\cal B}$ is high enough to easily reach diffraction efficiencies of more than 50\% for a few millimetre thick samples. Disadvantages are a rather difficult preparation procedure with combined thermo- and photo-polymerization, its profound ageing characteristics due to long lived radicals \cite{Fally-apb02,Havermeyer-00} and the expensive deuterated compounds. HPDLCs on the other hand are inexpensive, easy to prepare and manipulate. However, basically we tried only a single composition which is not optimised yet. Their major disadvantage for use as neutron optical elements is the limitation to thicknesses of about 50 micrometres. The reason for this limitation might be that light scattering from the LC component becomes dominating around this depth and thus the spatially modulated light pattern is destroyed. A workaround could be to record gratings at elevated temperatures and/or applied electric fields. For polymer-dispersed nanoparticles the main disadvantage again is the complicated preparation. The perspective of flexibility with respect to the incorporated particles is one major advantage, as well as the feasible thickness of about 200 $\mu$m \cite{Suzuki-ao07}. Various species and sizes of particles with peculiar properties open up possibilities to exploit interactions between neutrons and the solid state other than the nuclear one (magnetic interaction with ferromagnetic particles). A disadvantage which seems to be common to all materials probed up to now is their decreasing light-induced refractive index modulation, both for light and neutrons, with decreasing grating spacing.  We speculate that this is a property of acrylate based compounds. Therefore, a need to continue search for novel materials is evident.

We have demonstrated that H-PDLCs can be used as efficient diffraction elements for cold neutrons. An analysis of the accessible experimental parameters $\Lambda, d, \csld,\lambda$ showed that an optimised diffraction grating allowing for high diffraction efficiencies, Bragg diffraction regime and moderate angular selectivity could be realized with holographic polymer nanocomposites.  Most recent preliminary results let us anticipate that holographic polymer-dispersed nanoparticles \cite{Suzuki-ao07} would be the ideal candidates for versatile and efficient neutron optical gratings.
\ack
Financially supported by the Austrian Science Fund (P-18988 \& P-20265) and the \"OAD in the frame of the STC program Slovenia-Austria (SI-A4/0708). This research has been supported by the European Commission under the 6th Framework Programme through the Key Action: Strengthening the European Research Area, Research Infrastructures. Contract no.: RII3-CT-2003-505925.
We acknowledge continuous support by E. Tillmanns by making one of his labs available to us. We are grateful to M. Devetak  for providing the AFM graph.
\section*{References}

\end{document}